\documentclass[preprint2]{aastex}

\newcommand{\vt}{\mbox{\bf {T}}}
\newcommand{\vtcmb}{\mbox{\bf {T}}_{cmb}}
\newcommand{\vm}{\mbox{\bf {M}}} 
\newcommand{\vn}{\mbox{\bf {N}}} 
\newcommand{\vf}{\mbox{\bf {F}}} 
 
\input epsf 

\def\plotancho#1{\includegraphics[width=14cm]{#1}}

\begin{document}

\title{The Effect of Hot Gas in WMAP's First Year Data}

\author{C.  Hern\'andez--Monteagudo} \affil{Max Planck
Institut f\"ur Astrophysik (MPA).  \\ Karl-Schwarzschild
Str.  1.  D-85748 Garching bei M\"unchen, Germany}
\email{chm@mpa-garching.mpg.de}

\author{R.  Genova--Santos} \affil{Instituto de
Astrof\1sica de Canarias.  \\ C/ V\1a L\'actea, s/n.  38210
La Laguna, Tenerife, Spain\\} \email{rgs@ll.iac.es}

\and

\author{F.  Atrio--Barandela} \affil{F\1sica Te\'orica,
Universidad de Salamanca.\\ Plaza de la Merced s/n.  37008
Salamanca, Spain} \email{atrio@usal.es}

\begin{abstract}

By cross-correlating templates constructed from the 2 Micron
All Sky Survey (2MASS) Extended Source (XSC) catalogue with
WMAP's first year data, we search for the thermal
Sunyaev-Zel'dovich signature induced by hot gas in the local
Universe.  Assuming that galaxies trace the distribution of
hot gas, we select regions on the sky with the largest
projected density of galaxies. Under conservative
assumptions on the amplitude of foreground residuals, we
find a temperature decrement of -35 $\pm$ 7 $\mu$K
($\sim 5\sigma$ detection level, the highest reported
so far) in the $\sim$ 26 square
degrees of the sky containing the largest number of galaxies
per solid angle.  We show that most of the reported signal
is caused by known galaxy clusters which, when convolved with
the average beam of the WMAP W band channel, subtend a typical
angular size of 20--30 arcmins.  Finally, after removing
from our analyses all pixels associated with known optical and
X-ray galaxy clusters, we still find a tSZ decrement of -96
$\pm$ 37 $\mu$K in pixels subtending about $\sim$ 0.8 square
degrees on the sky.  Most of this signal is coming from five
different cluster candidates in the Zone of Avoidance (ZoA), present in the
Clusters In the ZoA (CIZA) catalogue.
We found no evidence that structures less bound than clusters
contribute to the tSZ signal present in the WMAP data.

\end{abstract}


\section{Introduction}

The study of the Cosmic Microwave Background (CMB) has
become a powerful cosmological tool with applications in
various astrophysical scenarios.  Recently, the WMAP team
has determined the main cosmological parameters from the CMB
temperature field with unprecedented accuracy
\citep{wmap_parm}.  This temperature field is composed of
primordial anisotropies, generated at the Last Scattering
Surface, and secondary fluctuations, which arise as the CMB
photons travel to the observer.  Among these secondary
anisotropies, we shall concentrate on the so called thermal
Sunyaev-Zel'dovich effect (hereafter tSZ, \citet{tSZ}),
associated with the distortion of the black body spectrum of
the CMB photons due to Compton scattering on fast moving
thermal electrons.  This spectral distortion is independent
of redshift and proportional to the integrated electron
pressure along the line of sight. It has been detected in
the direction of several galaxy clusters \citep{carlstrom02}. 
This makes the tSZ effect an useful
tool for the detection of ionized hot gas.

Recently, \citet{Fuk04} have argued that about 90\% of all
baryons are in the form of intergalactic plasma, not dense
and hot enough to be detectable as X-ray sources except in
clusters and groups of galaxies. At low redshfits, 30\% of all baryons
could be in the form of Ly-$\alpha$ absorbers
\citep{Penton04}, but a large fraction of them 
has not yet been accounted for observationally. 
Most models of structure
formation predict baryons to be located in filaments and
sheets, associated to galaxy overdensities
\citep{VSpringel}.  \citet{Xgas1} show evidence of
filamentary X-ray emission at the core of the Shapley
Supercluster, whereas \citet{Zappa02} report a detection of 
diffuse X-ray emission by warm gas ($T\sim 10^6$K)
associated with an overdense galaxy region.  The aim 
of this {\it letter} is to use the tSZ effect to detect
directly the diffuse warm baryon component in the local
Universe.  The analyses of the first year WMAP data have
indicated that a small contribution due to the tSZ induced by
clusters is present in the data (\citet{wmap_foreg},
\citet{scjal}).  \citet{pablo} and
\citet{afshordi} claimed  a tSZ detection at small angular
scales but did not clarify the nature of the astrophysical
sources associated with it.  \citet{Myers} 
analysed the angular extension of the tSZ signal in the W
band of WMAP by examining the CMB map in the direction of
the largest galaxy groups and clusters found in the catalogues of
\citet{aco} (hereafter ACO), \citet{apm} (hereafter APM)  and
the 2 Micron All Sky Survey Extended Source Catalogue, \citep{Jarrett}.
Their analyses showed evidence for
diffuse tSZ emission up to an angular scale of $\sim 1 \degr$,
which implied a baryon fraction larger than the WMAP estimate.

If baryons are distributed like the dark matter on
scales comparable to the virial radius of galaxies \citep{Fuk04},
the projected density of
galaxies on the sky will correlate with the tSZ signal,
{\em independently} of whether galaxies form clusters, groups or 
filaments. In this {\it letter}, we shall carry out a
pixel-to-pixel comparison of WMAP W band data with templates
constructed from the 2MASS galaxy catalogue, to test the
distribution of hot gas in the local Universe. In Sec.2 we
describe the pixel to pixel comparison method employed to estimate 
the tSZ signal and the data used in the analysis. In Sec.3 we present and
discuss our results, and we conclude in Sec.4.

\section{Method and Data Sets.}

The brightness temperature measured by a CMB experiment
is the sum of different components:
cosmological $\vtcmb$, tSZ, instrumental noise $\vn$ and
foreground residuals $\vf$.  If the tSZ signal is well
traced by a known spatial template, denoted here as $\vm$
and built for example from a galaxy catalogue, the total
anisotropy at a fixed position on the sky can be modelled as
$\vt = \vtcmb + {\tilde \alpha}\cdot\vm / \langle \vm \rangle + \vn + \vf$, 
where ${\tilde \alpha}$ measures the amplitude of the template induced
signal and $\langle \vm \rangle$ denotes the spatial average of the template.
 If all other components have zero mean and well
known correlation functions, then it is possible to use a
pixel to pixel comparison to estimate ${\tilde \alpha}$ (see
\citet{scjal} for details).  If ${\cal C}$ denotes the correlation
matrix of the CMB and noise components (foregrounds
residuals will be discussed later) then the estimate of
${\tilde \alpha}$ and its statistical error are
\begin{equation} 
\alpha = \frac{ \vt {\cal C}^{-1} \vm^T} {\vm {\cal C}^{-1} \vm^T }, 
\;\;\;\; \sigma_\alpha = \sqrt{\frac{1}{\vm{\cal C}^{-1}\vm^T}}.  
\label{eq:alpha1}
\end{equation} 
Since our galaxy template $\vm$ will be positive by construction, this 
equation demands that
CMB and noise fields have zero mean. We impose
the average of all pixels outside the Kp0 mask to be zero. We
checked that our results were insensitive to this requirement
by carrying out a similar analysis usig the more conservative Kp2
mask. In all cases our results changed by less than a few percent.
Notice that our method requires the inversion
of the correlation matrix, a computationally expensive
procedure.  To speed up the process we will carry out the
analysis in pixel subsets as described below.  We checked
with Monte Carlo simulations that $\sigma_\alpha$ is an
unbiased estimator of the error. 

We centered our analyses on WMAP W band, which has the
highest angular resolution.  In this band, the instrumental
noise shows almost no spatial correlation and its position
dependent amplitude is determined by the number of
observations\footnote{The WMAP data was downloaded from
{\it http://lambda.gsfc.nasa.gov}.}  \citep{wmap_noise}.
The tSZ template was built from the 2 Micron All Sky Survey
(2MASS) Extended Source Catalogue \citep{Jarrett}.  This
catalogue contains about 1.5 million galaxies, extending up
to redshift $z \sim 0.1$ (400 Mpc), detected in the
near-infrared ($J$, $H$ and $K_s$ bands). In order to
build a tSZ template ($\vm$), all 2MASS galaxies were projected
onto the sphere using the HEALPix\footnote{\it
http://www.eso.org/science/healpix/} pixelization
\citep{healpix} with the same resolution as the CMB data.
The amplitude in every pixel was made proportional to the 
number of galaxies. 
The template was then convolved with the window function of
the noise weighted average beam of the four Difference
Assemblies (DA's) corresponding to the WMAP W band (clean map),
and multiplied by the Kp0 mask, like the CMB data. Since our method
requires the CMB and noise components to have zero mean, we substracted
the map average outside the Kp0 mask.

Pixels were sorted in sets of size $N_{pix}$, denoted as
\vm$^{\beta}$, where the superscript index $\beta$ indicates galaxy
density, in such a way that 
low $\beta$ corresponds to higher projected galaxy density.
For patches of $N_{pix}=$2048, the average projected galaxy density
for $\beta=$1 was $\sim$ 420 galaxies per square degree, whereas for 
$\beta=$500 the density dropped to $\sim$ 44 galaxies per square degree,
which roughly coincides with the average projected density outside the
Kp0 mask. $N_{pix}$ ranged from 64 to 2048.  
The pixel to pixel comparison was then performed on
each of these subsets:  we compared {\it all} pixels in
\vm$^{\beta}$ to their counterparts in the CMB map. 
Our working hypothesis is that galaxies are fair tracers
of the gas density and within each set electron temperature
is similar, that is, in those pixels 
galaxies trace the electron pressure. 
Within each subset $\vm^{\beta}$, the galaxy density remained roughly
constant, so our method returned $\alpha$ as a weighted mean of
the measured temperature in those pixels. If no tSZ signal
is present, then $\alpha$ will scatter around zero, as a consequence
of CMB and noise being random fields of zero mean.

\section{Results and Discussion.}

\begin{figure*} 
\centering 
\plotancho{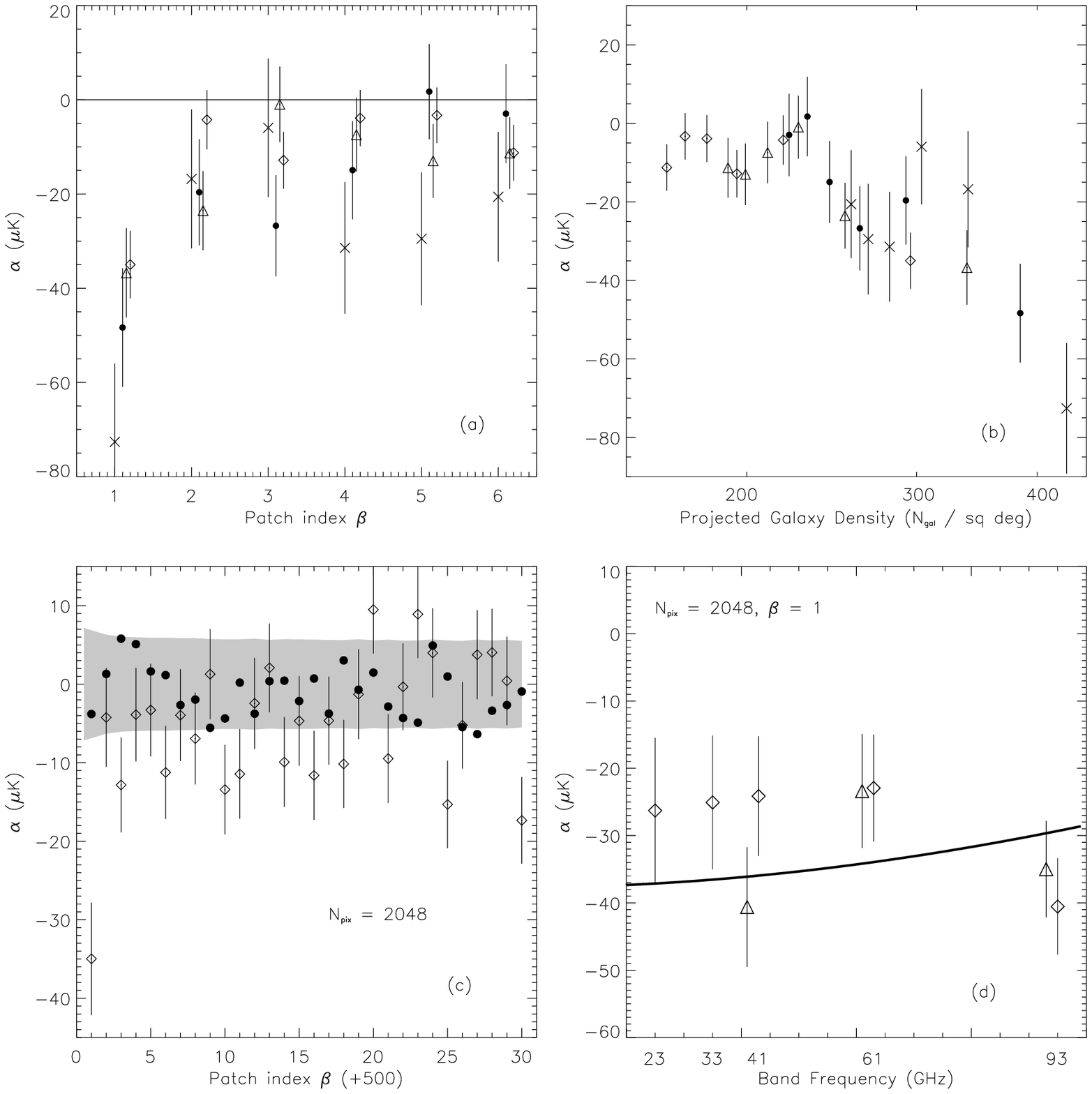} 
\caption[fig:allpxls]{{\it (a)} tSZ detection in pixel sets with the 
highest projected galaxy density.  Crosses, filled circles, triangles
and diamonds correspond to sets with $N_{pix} = 256, 512, 1024, 2048$,
respectively. Bars represent $1\sigma$ errors. {\it (b)} tSZ signal versus projected galaxy density.
The symbol coding is the same as in panel (a).
{\it (c)}  tSZ signal for
the very dense (\vm$^1$ to \vm$^{30}$, diamonds) and less dense 
(\vm$^{501}$ to \vm$^{530}$,
filled circles) sets. Bars and band are $1\sigma$ errors.
{\it (d)} tSZ for the different WMAP bands estimated on the densest set
of 2048 pixels. The solid line represents the expected frequency
dependence fitted to the {\em clean} Q, V and W band maps (triangles). 
Estimates of $\alpha$ obtained from raw (i.e., {\it non cleaned}) 
maps are displayed by diamonds. For bands with clean and raw maps,
triangles and diamonds are slightly shifted for better display.
} 
\label{fig:allpxls} 
\end{figure*}

Fig.~(\ref{fig:allpxls}) summarizes our main results:
Fig.~(\ref{fig:allpxls}a) shows the estimated $\alpha$'s for
the sets having the highest projected
density of galaxies.  In abscissas we give the set index
($\beta$).  Crosses, filled circles, triangles and diamonds
correspond to $N_{pix}=$ 256, 512, 1024 and 2048,
respectively.  Symbols are slightly shifted for
clarity, error bars denote 1$\sigma$ confidence levels.
At WMAP instrumental frequencies, the tSZ effect causes
temperature decrements, i.e., template and CMB data should 
anticorrelate giving negative $\alpha$'s, as found.  
Sets 1 and 2 of size $N_{pix}=256$ correspond to the
first set of $N_{pix}=512$ and similarly for all other sizes
and sets.  Consistently, $\alpha$ of a larger set is always
bracketed by the $\alpha$'s measured from its subsets.  The
largest signal comes from the densest 256 pixels but the
highest statistical level of significance is achieved for
$N_{pix}=2048$ ($\alpha = -35\;\mu$K at the $4.9\sigma$
detection level), since the error bars shrink due to the
higher number of pixels contributing with tSZ signal.
In Fig.(\ref{fig:allpxls}b) the same data sets are plotted versus 
the average projected galaxy density, i.e., the
average number of galaxies per square degree within each subset, as seen
by the W band beam. The symbol coding is identical: first sets of
$N_{pix}=$ 256, 512 show a projected galaxy density as high as
420--380 galaxies per square degree, whereas the 6th set for
$N_{pix}=$ 2048 contains around 140 galaxies per square degree.

In order to evaluate the significance of the previous
results, we repeat the analysis for pixels of intermediate
and low projected density of galaxies.  In
Fig.~(\ref{fig:allpxls}c) diamonds correspond to sets of
size $N_{pix}=2048$ with $\beta \in [1,30]$, whereas filled
circles correspond to indices $\beta \in [501, 530]$.  The
shaded area limits the 1$\sigma$ error bar for diamonds,
that is about a factor of 1.2 bigger than for filled
circles.  While the latter scatter around zero with the
expected dispersion, diamonds are clearly biased towards
negative values: besides the first patch ($\alpha = -35
\pm 7\;\mu$K), there are {\em seven} other sets above 
the 2$\sigma$ level, one of them at $3\sigma$.  

We also tested the consistency of our results with respect
to frequency. In Fig.~(\ref{fig:allpxls}d) we estimate $\alpha$ by
cross-correlating the densest pixel set
($N_{pix}=2048$, $\beta=1$) with all WMAP bands:  K (23GHz),
Ka (33GHz), Q (41GHz), V(61GHz) and W(93GHz). Diamonds give the
obtained $\alpha$'s from {\it raw} maps, whereas triangles refer to analyses
performed on {\em foreground cleaned} maps available in the LAMBDA
site. Note that the
WMAP team only provided clean maps
for the three highest frequency channels. In all maps, the
signal is compatible with being due to tSZ.  A more
quantitative comparison is not straighforward since the maps
have different angular resolution and galactic contamination.  When
compared in pairs, all $\alpha$'s are within 1$\sigma$ of
each other and they are all within 2$\sigma$ of the expected
frequency dependence of the tSZ effect, whose best fit
for the cleaned Q, V and W band maps is plotted as a solid line.

\begin{figure}[h] 
\begin{center} \epsfxsize=6.cm \epsfbox{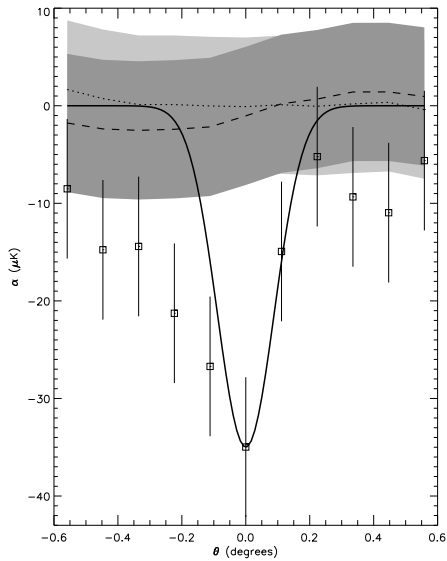} 
\caption[fig:rottest]{Rotation test showing the angular extent of the reported
tSZ sources: $\alpha$ is computed for different angular displacements of
the galaxy template with respect to the {\it cleaned} W band map, 
for the densest 2048
pixels (squares, error bars at 1$\sigma$). This analysis is repeated on 
100 MC simulations including CMB, W-band noise {\it plus} (dotted line) and 
{\it minus} (dashed line) our foreground template. Shaded areas denote again
1$\sigma$ confidence level. The solid lines displays the gaussian
approximation to the W band beam.}
\label{fig:rottest}
\end{center} 
\end{figure}

We measure the extent of the tSZ regions by rotating the
template around the z-axis (perpendicular to the galactic plane)
and cross-correlating the densest 2048 pixels.
In  Fig.~(\ref{fig:rottest}) squares give $\alpha$
versus the average angular displacement of the pixel set.
Error bars are again 1$\sigma$. The solid line represents 
the gaussian approximation of the W-band beam.
The size of the tSZ sources is typically 20--30 arcmins, slightly
bigger than the beam but remarkably smaller than
the values obtained by \citet{Myers} from rich ACO clusters.
Actually, when one studies the angular distribution of the 2048 densest
pixels, one finds that they are associated in groups of typically 3--4
members, and that the groups are uniformly distributed on the sky.
 
By comparing, in Fig.~(\ref{fig:allpxls}d), results from the
raw and {\em cleaned} maps built from the W band,
 one can conclude that foregrounds
will have little impact on our results. We made a
more detailed study using Monte Carlo simulations:
we performed
100 simulations of the CMB and W-band noise components and added
and removed a foreground residual template. As a conservative model 
for foreground residuals left after cleaning the W-band, we 
took the sum of the dust, free-free and synchrotron emission
maps for the W band released by the WMAP team. In each simulation
we (i) added and (ii) substracted {\it the whole foreground template}.
In  Fig.~(\ref{fig:rottest}) we plot the average $\alpha$'s for 
these simulations, after adding (dotted line)  and substracting (dashed line) 
the residuals for several angular displacements.
The shaded bands display the 1$\sigma$ dispersion areas. 
The errors practically equal the statistical estimates of 
eq.~(\ref{eq:alpha1}). To summarize,
foregrounds do not significantly affect our estimated amplitude of 
the tSZ contribution.

\begin{figure}[h] 
\begin{center} 
\epsfxsize=8.cm \epsfbox{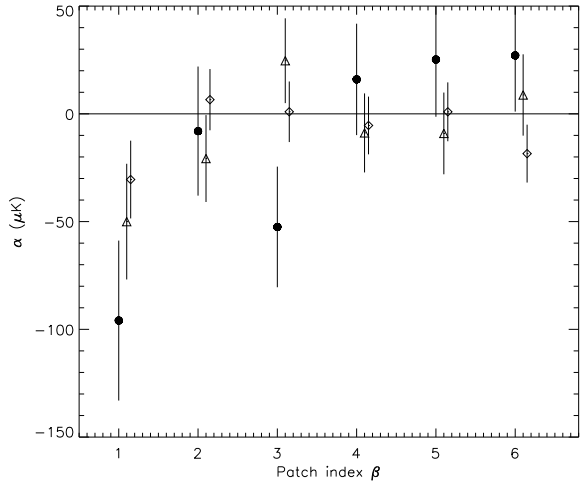}
\caption[fig:diffuse]{Persistence of tSZ signal once all
known clusters have been excised from the analysis. Filled circles,
triangles and diamonds correspond to patches of $N_{pix}=$ 64, 128 and 256,
 respectively.}
\label{fig:diffuse} 
\end{center} 
\end{figure}

\begin{figure}[h] 
\begin{center} 
\epsfxsize=10.1cm \epsfbox{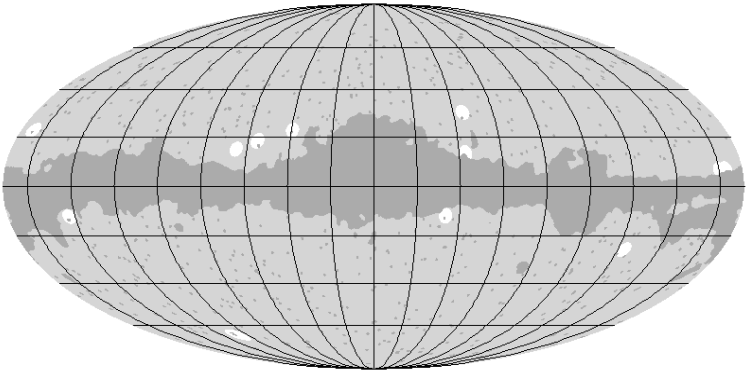}
\caption[fig:mappxls]{White spots show the position of the 64 densest
pixels not excised after removing
known galaxy clusters. These pixels contribute with a tSZ signal
of amplitude of $\alpha = -96 \pm 37\; \mu$K. The dark grey
area is masked out by the Kp0 mask, which covers the Galaxy and the
brightest radio sources. The graticule is 20\degr $\times$ 20\degr.
}
\label{fig:mappxls} 
\end{center} 
\end{figure}

Finally, we removed from our galaxy template all those pixels
that were associated with known galaxy clusters.  We used the
ACO and APM catalogues of optically
selected clusters and the XBC
\citep{Ebeling00},
de Grandi \citep{deGrandi}, NORAS \citep{Noras}, ROSAT-PSPC
\citep{pspc} and Vogues \citep{Vogues} X-ray cluster
catalogues.  We excised from the analyses all pixels lying
within a virial radius of the cluster center (taken to be ten times 
the core radius). For clusters without measured core
radius but with known redshift, we assumed a virial radius
of 1.7 Mpc, and removed all pixels within that
distance.  For the rest (the
majority), we conservatively removed all pixels within a
circle of 30 arcmin from the cluster centre.  
Out of the 2048 pixels for $\beta=1$, 1681 were eliminated. 
For the patches with $\beta = 2-30$, i.e., the next 
$\sim$60,000 densest pixels 
in Fig.(\ref{fig:allpxls}c), a large fraction of them were also
associated to known clusters and were eliminated with the excising.
In Fig.(\ref{fig:diffuse}) we show the cross correlation of
these remaining pixels outside known clusters with the clean W band map.
Note that here patch sortening has been regenerated using surviving pixels.
For the densest sets of $N_{pix}=$ 64 (filled circles), 128 (triangles)
and 256 (diamonds), we still find evidence of tSZ, but at
much lower level of significance.  For the densest 64 pixels,
subtending $\sim 0.8$ square degrees on the sky, we
obtain $\alpha=-96\pm 37\;\mu$K, at $\sim 2.6\sigma$
significance level.  The signal gets dilluted rapidly as more
pixels are included in the analysis:  $\alpha = -50 \pm 27\;
\mu$K and $\alpha = -30 \pm 18\; \mu$K for $N_{pix}=$ 128,
256, respectively. Out of the 64 densest pixels, 54 pixels
are in the ZoA, and 45 of them coincide
with five different cluster candidates in the CIZA \citep{ciza} catalogue,
(Ebeling, private communication).
The remaining group of pixels are not associated to any known 
galaxy cluster. In Fig.~(\ref{fig:mappxls})
we plot the location of those pixels in the sky.
The shaded area correspond to the Kp0 mask, that also
remove many point sources outside the galactic plane
(dark grey dots). The 64 pixels are ploted as big white
circles for convenience.

\section{Conclusions}

Under the assumption that galaxies trace hot gas, we have used
the 2MASS galaxy catalogue to
search for tSZ signal present in WMAP data.
In $\sim 26$ square degrees on the sky we have
found a contribution of average amplitude -35 $\pm$ 7 $\mu$K, 
spectrally compatible with tSZ, and mostly
generated by ACO clusters of galaxies.  
Our study, based on a pixel-to-pixel comparison, reaches the highest
sensitivity level reported so far.
Compared with methods based on power spectrum or correlation function
analysis, our method gives a larger level of significance
since we restrict the 
analysis to the regions of the sky where the tSZ contribution 
is expected to be the largest.  

We have found that
the typical angular extension of this signal is somewhere
between 20--30 arcmins.  Furthermore, once all known
clusters of galaxies are excised from the analysis, we are
left with $\sim 0.8$ square degrees with an average amplitude of
$\alpha = -96 \pm 37\; \mu$K:  those pixels fall mostly in 
the ZoA and after performing our analyses
we found that 45 of them are associated to five different galaxy
clusters in the CIZA catalogue. 
We have found no conclusive evidence that, 
in the volume probed by 2MASS, structures less
bound than clusters contribute to the tSZ signal present in the WMAP data.

\acknowledgments We thank R.Rebolo and J.A.Rubi\~no--Mart\1n
for enlightening discussions. We also thank H.Ebeling for comments on
the CIZA catalogue and an anonymous referee for useful criticism.
C.H.M. acknowledges the financial support
from the European Community through the Human Potential
Programme under contract HPRN-CT-2002-00124 (CMBNET)
and useful discussions with V.M\"uller, R.Croft and
A.Banday. F.A.B. acknowledges finantial support
from the Spanish Ministerio de Educaci\'on y Ciencia
(projects BFM2000-1322 and AYA2000-2465-E)
and from the Junta de Castilla y Le\'on (project SA002/03).
Some of the results in this paper have been derived using the HEALPix
package, \citep{healpix}.
We acknowledge the use of the Legacy Archive for Microwave
Background Data Analysis (LAMBDA, http://lambda.gsfc.nasa.gov).
Support for LAMBDA is provided by the NASA Office of Space Science.
This publication makes use of data products from the Two Micron All Sky
Survey, which is a joint project of the University of Massachusetts and
the Infrared Processing and Analysis Center/California Institute of
Technology, funded by the National Aeronautics and Space Administration
and the National Science Foundation.

\clearpage

\clearpage

\clearpage

\end{document}